\documentclass[10pt,a4paper,onecolumn]{IEEEtran}
\usepackage[lmargin=2cm, rmargin=2cm, tmargin=2cm, bmargin=2cm]{geometry}
\usepackage{ccaption}
\usepackage[T1]{fontenc}
\usepackage{graphicx}
\usepackage{fancyhdr}
\usepackage{helvet}
\usepackage{floatflt}
\usepackage{colortbl}
\usepackage{fancyvrb}
\usepackage{wrapfig}
\usepackage[latin1]{inputenc}
\usepackage[T1]{fontenc}
\usepackage[english]{babel}
\usepackage{graphicx}
\usepackage{subfigure}
\usepackage{amsfonts}
\usepackage{amsmath}
\usepackage{theorem}
\usepackage[round,authoryear]{natbib}
\usepackage{ae}
\usepackage{bm}

\newtheorem{assumption}{Assumption}

\DeclareMathOperator{\sgn}{sgn}

\title{An adaptive fuzzy sliding mode controller applied to a chaotic pendulum}
\author{Wallace Moreira Bessa, Aline Souza de Paula, Marcelo Amorim Savi}
\date{}

\begin{document}

\maketitle

\abstract{

In this work, an intelligent controller is employed to the chaos control problem in a nonlinear pendulum. 
The adopted approach is based on the sliding mode control strategy and enhanced by an adaptive fuzzy algorithm 
to cope with modeling inaccuracies. The convergence properties of the closed-loop system are analytically 
proven using Lyapunov's direct method and Barbalat's lemma. Numerical results are also presented in 
order to demonstrate the control system performance.

}

\section*{Introduction}

Chaotic response is related to a dense set of unstable periodic orbits (UPOs) and the system often 
visits the neighborhood of each one of them. Moreover, chaos has sensitive dependence to initial 
condition, which implies that the system evolution may be altered by small perturbations. Chaos 
control is based on the richness of chaotic behavior and may be understood as the use of tiny 
perturbations for the stabilization of an UPO embedded in a chaotic attractor.  It makes this 
kind of behavior to be desirable in a variety of applications, since one of these UPO can provide 
better performance than others in a particular situation. Due to these characteristics, chaos and 
many regulatory mechanisms control the dynamics of living systems. Inspired by nature, it is possible 
to imagine situations where chaos control is employed to stabilize desirable behaviors of mechanical 
systems. Under this condition, these systems would present a great flexibility when controlled, being 
able to quickly change from one kind of response to another. Literature presents some contributions 
related to the analysis of chaos control in mechanical systems. \citet{andrievskii1} and \citet{savi3} 
present an overview of applications of chaos control in various scientific fields. 

Intelligent control, on the other hand, has proven to be a very attractive approach to cope with 
uncertain nonlinear systems 
\citep{tese,cobem2005,Bessa2017,Bessa2018,Bessa2019,Deodato2019,Lima2018,Lima2020,Lima2021,Tanaka2013}. 
By combining nonlinear control techniques, such as feedback linearization or sliding modes, with 
adaptive intelligent algorithms, for example fuzzy logic or artificial neural networks, the resulting 
intelligent control strategies can deal with the nonlinear characteristics as well as with modeling 
imprecisions and external disturbances that can arise.

This contribution proposes a robust intelligen controller that can be applied to stabilize 
UPOs of chaotic attractors. The adopted approach is based on the sliding mode control strategy 
and enhanced by a stable adaptive fuzzy inference system to cope with modeling inaccuracies and 
external disturbances that can arise. The boundedness of all closed-loop signals and the 
convergence properties of the tracking error are analytically proven using Lyapunov's direct 
method and Barbalat's lemma. The general procedure is applied to a nonlinear pendulum that 
presents chaotic response \citep{paula1}. Numerical simulations are carried out showing the 
stabilization of some UPOs of the chaotic attractor showing an effective response, demonstrating 
the controller performance.

\section*{Chaotic pendulum}

The nonlinear pendulum consists of an aluminum disc with a lumped mass that is connected to a rotary 
motion sensor. This assembly is driven by a string-spring device that is attached to an electric motor 
and also provides torsional stiffness to the system. A magnetic device provides an adjustable dissipation 
of energy. An actuator provides the necessary perturbations to stabilize this system by properly changing 
the string length. It is assumed that system dissipation may be expressed by a combination of a linear 
viscous dissipation together with dry friction. Therefore, denoting the angular position as $\phi$, the 
following equation is obtained.

\begin{equation}
\ddot{\phi}+\frac{\zeta}{I}\dot{\phi}+\frac{kd^2}{2I}\phi+\frac{\mu\sgn(\dot{\phi})}{I}+\frac{mgD
\sin(\phi)}{2I}=\frac{kd}{2I}\left(\sqrt{a^2+b^2-2ab\cos(\omega t)}-(a-b)-\Delta l\right)
\label{eq:pend1}
\end{equation}

\noindent
where $\omega$, $a$, $b$, $D$, $d$, $m$, $\zeta$, $\mu$, $I$, $k$ are system parameters; $g$ is the 
gravity acceleration and $\sgn(x)$ is the sign of the variable $x$. \citet{paula1} show that this 
mathematical model is in close agreement with experimental data and, therefore, it will be used for 
the control purposes.

\section*{Adaptive fuzzy sliding mode control}

In order to write Eq.~(\ref{eq:pend1}) in a more convenient form, it is rewritten as follows:

\begin{equation}
\ddot{\phi}=f(\dot{\phi},\phi,t)+hu+p
\label{eq:pend2}
\end{equation}

\noindent
where $h=kd/2I$, $u=-\Delta l$, $f$ can be obtained from Eq.~(\ref{eq:pend1}) and Eq.~(\ref{eq:pend2}),
and the added term $p$ represents both unmodeled dynamics and external disturbances.

Now, let $S(t)$ be a sliding surface defined in the state space by the equation $s(\dot{e},e)=0$, 
with the function $s:\mathbb{R}^2\to\mathbb{R}$ satisfying $s(\dot{e},e)=\dot{e}+\lambda e$, where 
$e=\phi-\phi_d$ is the tracking error, $\dot{e}$ is the first time derivative of $e$, $\phi_d$ is the 
desired trajectory and $\lambda$ is a strictly positive constant. The controlling of the system dynamics 
(\ref{eq:pend2}) is done by assuming a sliding mode based approach, defining a control law composed by 
an equivalent control $\hat{u}=\hat{h}^{-1}(-\hat{f}-\hat{p}+\ddot{\phi}_d-\lambda\dot{e})$ and a 
discontinuous term $-K\sgn(s)$:

\begin{equation}
u=\hat{h}^{-1}(-\hat{f}-\hat{p}+\ddot{\phi}_d-\lambda\dot{e})-K\sgn(s)
\label{eq:usgn}
\end{equation}

\noindent
where $\hat{h}$, $\hat{f}$, and $\hat{p}$ are estimates of $h$, $f$ and $p$, respectively, and
$K$ is a positive control gain.

Regarding the development of the control law, the following assumptions should be made:

\begin{assumption}
The function $f$ is unknown but bounded, i.e. $|\hat{f}-f|\le\mathcal{F}$.
\label{hp:limf}
\end{assumption}
\begin{assumption}
The input gain $h$ is unknown but positive and bounded, i.e. $0<h_{\mathrm{min}}\le h\le 
b_{\mathrm{max}}$.
\label{hp:limh}
\end{assumption}
\begin{assumption}
The perturbation $p(t)$ is time-varying and unknown but bounded, i.e. $|p(t)|\le\mathcal{P}$.
\label{hp:limp}
\end{assumption}

Based on Assumption~\ref{hp:limh} and considering that the estimate $\hat{h}$ could be chosen according
to the geometric mean $\hat{h}=\sqrt{h_\mathrm{max}h_\mathrm{min}}$, the bounds of $h$ may be expressed
as $\mathcal{H}^{-1}\le\hat{h}/h\le\mathcal{H}$, where $\mathcal{H}=\sqrt{h_\mathrm{max}/h_\mathrm{min}}$.

Under this condition, the gain $K$ should be chosen according to

\begin{equation}
K\ge\mathcal{H}\hat{h}^{-1}(\eta+|\hat{p}|+\mathcal{P}+\mathcal{F})+(\mathcal{H}-1)|\hat{u}|
\label{eq:gain}
\end{equation}

\noindent
where $\eta$ is a strictly positive constant related to the reaching time.

At this point, it should be highlighted that the control law (\ref{eq:usgn}), together with (\ref{eq:gain}), 
is sufficient to impose the sliding condition

\begin{equation}
\frac{1}{2}\frac{d}{dt}s^2\le-\eta|s|
\label{eq:slidcond}
\end{equation}

\noindent
and, consequently, the finite time convergence to the sliding surface $S$.

In order to obtain a good approximation to the disturbance $p(t)$, the estimate $\hat{p}$ will be 
computed directly by an adaptive fuzzy algorithm. The adopted fuzzy inference system was the zero 
order TSK (Takagi--Sugeno--Kang), whose rules can be stated in an appropriate linguistic manner. 
Considering that each rule defines a numerical value as output $\hat{P}_r$, the final output 
$\hat{p}$ can be computed by the dot product:

\begin{equation}
\hat{p}(s) = \mathbf{\hat{P}}^{\mathrm{T}}\mathbf{\Psi}(s)
\label{eq:dcvet}
\end{equation}

\noindent
where $\mathbf{\hat{P}}=[\hat{P}_1, \hat{P}_2, \dots, \hat{P}_N]^{\mathrm{T}}$ is the vector containing 
the attributed values $\hat{P}_r$ to each rule $r$, $\mathbf{\Psi}(s)=[\psi_1(s), \psi_2(s),$ $\dots, 
\psi_N(s)]^{\mathrm{T}}$ is a vector with components $\psi_r(s)= w_r/\sum_{r=1}^{N}w_r$ and $w_r$ is 
the firing strength of each rule. In order to obtain the better estimation $\hat{p}(s)$ to the disturbance 
$p$, the vector of adjustable parameters can be automatically updated by $\mathbf{\dot{\hat{P}}}=\varphi 
s \mathbf{\Psi}(s)$, where $\varphi$ is a strictly positive constant related to the adaptation rate. 

In order to evaluate the stability of the closed-loop system, let a positive-definite function 
$V$ be defined as

\begin{equation}
V(t)=\frac{1}{2}s^2+\frac{1}{2\varphi}\bm{\delta}^\mathrm{T}\bm{\delta}
\label{eq:liap}
\end{equation}

\noindent
where $\bm{\delta}=\mathbf{\hat{P}}-\mathbf{\hat{P}}^*$ and $\mathbf{\hat{P}}^*$ is the optimal
parameter vector, associated with the optimal estimate $\hat{p}^*(s)$. Thus, by considering the 
time derivative of $V$ and defining a minimum approximation error as $\varepsilon=\hat{p}^*(s)-p$, 
and recalling the definitions of $s$, $u$ and $\dot{\hat{P}}$, it is possible to verify that 
$\dot{V}$ becomes

\begin{equation}
\dot{V}(t)=-\Big[(\hat{f}-f)+\varepsilon+\hat{h}^{-1}\hat{u}-h\hat{u}-hK\mbox{sgn}(s)\Big]s\le-\eta|s|
\label{eq:liapdot}
\end{equation}

\noindent
Here assumptions~\ref{hp:limf}--\ref{hp:limp} are evoked and $K$ is defined according to (\ref{eq:gain}). 
This implies $V(t)\le V(0)$ and that $s$ and $\bm{\delta}$ are bounded. Integrating both sides of 
(\ref{eq:liapdot}) and evoking Barbalat's lemma it is established that $s\to0$ as $t\to\infty$, which 
ensures the convergence of the states to the sliding surface $S$ and to the desired trajectory. At this 
point, it should be notice that discontinuous terms can produce undesirable high frequency oscillations 
of the controlled variable. Therefore, it is convenient to use saturation functions that smoothes system 
discontinuities \citep{slotine1}. 

\section*{Numerical simulations}

\begin{figure}[htb]
\centering
\includegraphics[width=0.4\textwidth]{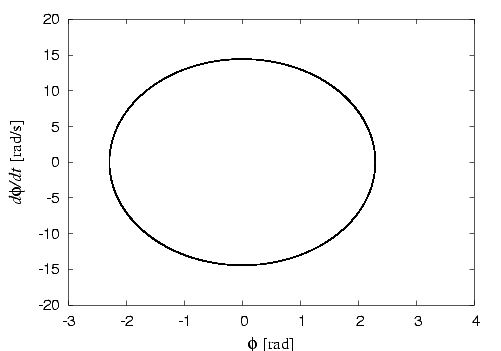}
\includegraphics[width=0.4\textwidth]{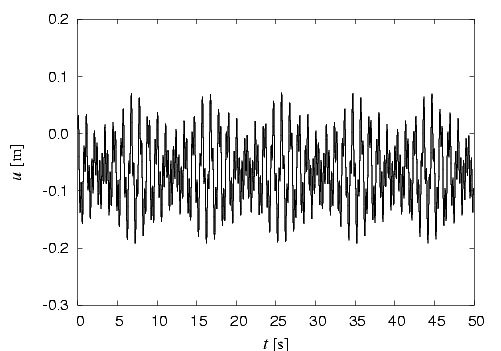}
\caption{Tracking of $[\dot{\phi}_d,\phi_d]=[4.6\pi\cos(2\pi t),2.3\sin(2\pi t)]$.}
\label{fg:sim1}
\end{figure}

\begin{figure}[htb]
\centering
\includegraphics[width=0.4\textwidth]{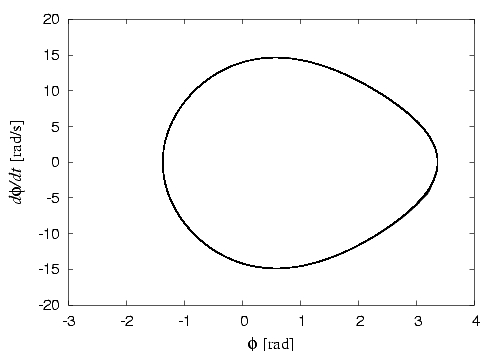}
\includegraphics[width=0.4\textwidth]{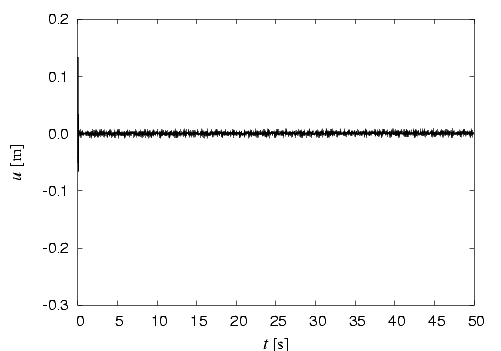}
\caption{Tracking of a Period-1 UPO.}
\label{fg:sim2}
\end{figure}

In order to analyze the controller perfomance, numerical simulatations are carried out considering the 
fourth order Runge-Kutta method. The model parameters are chosen according to \citet{paula1} and control 
parameters are $\lambda=10.0$, $\eta=0.5$, $\gamma=1.0$ and $\mathcal{H}=1$. Basically, two different 
situations are treated. In the first case, Figure~\ref{fg:sim1}, a generic orbit $[\dot{\phi}_d,\phi_d]
=[4.6\pi\cos(2\pi t),2.3\sin(2\pi t)]$ are considered, while in the second case, Figure~\ref{fg:sim2}, 
a period-1 UPO are chosen. Although both orbits are similar, it should be highlighted that the control 
action $u$ the controller requires less effort to stabilize an UPO.

\section*{Concluding remarks}

The present contribution considers the stabilization of orbits employing an adaptive fuzzy sliding 
mode controller. The stability and convergence properties of the closed-loop systems is proven using 
Lyapunov stability theory and Barbalat's lemma. As an application of the general formulation, numerical 
simulations of a nonlinear pendulum with chaotic response is of concern. The control system performance 
is investigated showing for the tracking of a generic orbit as well as for UPO stabilization. It is 
shown that the controller needs less effort to stabilize an UPO. 


\end{document}